\documentclass[useAMS,usenatbib]{mn2e} \input{epsf}

\newif\ifAMStwofonts
\AMStwofontstrue

\usepackage{ulem} 
\usepackage{longtable}
\usepackage{lscape}
\usepackage{float}
\usepackage{url}
\usepackage[english]{babel}
\usepackage[utf8]{inputenc}
\usepackage[varg]{txfonts}
\usepackage{graphicx}
\usepackage{txfonts}
\usepackage{natbib}
\usepackage[title]{appendix}
\usepackage[nameinlink]{cleveref}

\makeatletter
\newlength{\abovecaptionskip}
\makeatother
\usepackage{threeparttable}
\usepackage{xcolor}
\usepackage{ulem}

\def\lesssim{\mathrel{\hbox{\rlap{\hbox{\lower4pt\hbox{$\sim$}}}\hbox{$<$}}}}
\def\gtrsim{\mathrel{\hbox{\rlap{\hbox{\lower4pt\hbox{$\sim$}}}\hbox{$>$}}}}

\def\l_lsun{$\log{L/\rm L_{\odot}}$~}
\def\masa_msun{$M/ \rm M_{\odot}$~}

\def\m_mstar{$M/M_{*}$~}

\title[Tides on the orbit of redback PSR~J1723-2837]{Effect of tides on the orbital evolution of the  redback system PSR~1723-2837}
 \author[M. L. Novarino, M. Echeveste, O. G. Benvenuto, M. A. De Vito \& G.A. Ferrero]
 {
 M. L. Novarino\thanks{Fellow of the Consejo Nacional de Investigaciones Cient\'{\i}ficas y T\'ecnicas (CONICET). Emails: leonova@fcaglp.unlp.edu.ar, mecheveste@fcaglp.unlp.edu.ar},
 M. Echeveste$^{\star}$,
 O. G. Benvenuto\thanks{Member of  the Carrera del Investigador
 Cient\'{\i}fico, Comisi\'on de  Investigaciones Cient\'{\i}ficas de la  Provincia de Buenos Aires (CIC). Email: \mbox{obenvenu@fcaglp.unlp.edu.ar}},
 M. A. De  Vito\thanks{Member of  the Carrera del Investigador
 Cient\'{\i}fico of CONICET. Email: adevito@fcaglp.unlp.edu.ar}
 and G.A. Ferrero 
 \\
  Instituto de Astrof\'{\i}sica de La Plata, IALP, CCT-CONICET-UNLP, Argentina and\\
 Facultad de Ciencias Astron\'omicas y Geof\'{\i}sicas, Universidad
 Nacional de La Plata (UNLP),\\ Paseo del Bosque S/N, B1900FWA, La Plata,
 Argentina}

\begin{document}


\pagerange{\pageref{firstpage}--\pageref{lastpage}} \pubyear{2020}

\maketitle \label{firstpage}

\begin{abstract}

The standard model of stellar evolution in Close Binary Systems assumes that during mass transfer episodes the system is in a synchronised and circularised state. Remarkably, the redback system PSR~J1723-2837 has an orbital period derivative $\dot{P}_{orb}$ too large to be explained by this model. Motivated by this fact, we investigate the action of tidal forces in between two consecutive mass transfer episodes for a system under irradiation feedback, which is a plausible progenitor for PSR~J1723-2837. We base our analysis on Hut's treatment of  equilibrium tidal evolution, generalised by considering the donor as a two layers object that may not rotate as a rigid body. We also analyse three different relations for the viscosity with the tidal forcing frequency. We found that the large value measured for $\dot{P}_{orb}$ can be reached by systems where the donor star rotates  slower (by few percent) than the orbit just after mass transfer episodes. Van Staden \& Antoniadis have observed this object and reported a lack of synchronism, opposite to that required by the Hut's theory to account for the observed $\dot{P}_{orb}$. Motivated by this discrepancy, we analyse photometric data obtained by the spacecraft Kepler second mission K2, with the purpose of identifying the periods present in PSR~J1723-2837. We notice several periods close to those of the orbit and the rotation. The obtained periods pattern reveals the presence of a more complex phenomenology, which would not be well described in the frame of the weak friction model of equilibrium tides.   

\end{abstract}


\section{Introduction} \label{sec:intro}

Redbacks are eclipsing binary systems composed by a neutron star (NS) and a non-degenerate, low mass companion ($0.1 < M_{2}/M_{\odot}< 0.7$) with an orbital period between 0.1 and 1 day. According to the standard scenario (see, for example, \citealt{2002ApJ...565.1107P}), these systems are a result of the evolution of interacting binaries in which the stars exchange mass. In systems that may be redbacks progenitors, the donor is a solar-like star that evolves as it were isolated until it eventually fills its Roche Lobe, allowing matter and angular momentum to flow towards the NS. The standard model predicts a long and stable episode of mass transfer and, sometimes, a small number of Roche Lobe Overflows (RLOFs) due to thermonuclear flashes of the donor. Adding irradiation feedback to this model is relevant when studying systems with short orbital periods, such as redbacks \citep{2011Sci...333.1717B}. Irradiation feedback occurs when the donor star transfers mass onto the NS. The transferred matter produces X-ray radiation that illuminates the donor star; if this star has an outer convective zone, the irradiated surface is partially inhibited from releasing energy emerging from its deep interior. In some cases, the star's structure is unable to sustain the RLOF and the donor detaches. Then, nuclear evolution may lead the donor star to experience a RLOF again, and the X-ray radiation from the matter that falls onto the NS reappears. As a consequence of irradiation feedback, systems suffer several pulses of mass transfer (see, for example, \citealt{1993A&A...277...81H}; \citealt{2004A&A...423..281B}; \citealt{2012ApJ...753L..33B}). The number of mass transfer pulses depends on a free parameter of the model, called $\alpha_{irrad}$, which represents the fraction of the incident flux that effectively irradiates the donor star. Larger values of this parameter would result in a smaller number of RLOFs. Irradiated systems can be detected as X-ray sources during RLOF states, or as binary millisecond pulsars once the mass transfer ceases. 

Besides, in the standard scenario, when studying systems which descend from interacting binaries, it is usually supposed that the systems are tidally locked, so they have a synchronous rotation, a circular orbit, and the rotational axes of both stars are aligned with the orbital rotation axis. This implies that the orbital period of the binary equals the rotational period of the donor star. These assumptions may be useful to study the general properties of the systems, but it may result inadequate for investigating the changes in the orbital period. 

In the frame of the model presented above, \citet{2015ApJ...798...44B} found a plausible progenitor to the redback system PSR~J1723-2837 that accounts for its main characteristics (orbital period, mass, mass ratio, and temperature of the donor star), given by \citet{2004MNRAS.355..147F} and \citet{2013ApJ...776...20C}. This progenitor is in a low-irradiation regime with $\alpha_{irrad}=~0.01$. Systems with greater values of $\alpha_{irrad}$ are not plausible progenitors since at the time they have the mass ratio observed ($M_{NS} / M_2 = 3.3\pm0.5$), they are transferring mass, so they would be observed as low-mass X-ray binaries (LMXBs), see Sec.~4 in \cite{2015ApJ...798...44B}. Remarkably, the temporal derivative of the orbital period predicted by the model is three orders of magnitude smaller than the observed value, $\dot{P}_{orb}=~-3.50(12) \times 10^{-9} ss^{-1}$ \citep{2013ApJ...776...20C}. This observed value is large in comparison with the values of $\dot{P}_{orb}$ measured in other close binary systems (CBS), as can be seen in Table~\ref{tab:Ppunto}.  We consider that a so large disparity between observations and theoretical predictions may indicate the necessity of improving the models. It is the aim of this paper to explore the physical processes that can lead the PSR~J1723-2837 system to have a so large negative value of the temporal derivative of the orbital period $\dot{P}_{orb}$.

The disparity between observations and theory for this redback motivated us to study tidal interactions in CBSs. In systems like redbacks, the compact object introduces a differential force that raises tides on the surface of the donor. If there were not dissipation of kinetic energy into heat, the star would elongate just in the direction of the line joining the centres of mass of both stars. Instead, that dissipation induces a phase lag in the tidal bulge, so the inclined mass portion exerts a torque on the star. This torque leads to an exchange of angular momentum between the stellar spin and the orbit, while conserving the total angular momentum and diminishing the orbital and rotational energy. In consequence, the orbital parameters change, and stellar rotation tends to synchronise with the orbital motion, the orbit tends to circularise and the equatorial plane approaches the orbital plane (\citealt{1981A&A....99..126H,2008EAS....29...67Z, 2014MNRAS.444..542R}). During the synchronisation process, the orbital period changes, allowing the occurrence of large values of $\dot{P}_{orb}$. 

Here we study the effect of tidal forces acting on the donor star in the redback system PSR~J1723-2837. We consider the models presented in \citet{2015ApJ...798...44B}, in which they found a progenitor for this redback. In that paper, it was assumed that the system is always synchronised. Here we shall relax this hypothesis. As the derived observed radius of the donor star indicates that it is close to filling its Roche Lobe, we explore the effects of tides in between two consecutive mass transfer episodes, i.e., when the system is in a quasi-RLOF state. We work under the weak friction approximation in the equilibrium tide. Equilibrium means that the star is assumed to be in hydrostatic equilibrium and that, if there were no dissipation mechanisms, it would instantaneously adjust to the perturbing force exerted by its companion, i.e., the NS. The weak friction model supposes that the tidal lag angle, produced by the phase lag in the tidal bulge, is proportional to the difference between the orbital angular velocity and the rotational velocity of the star \citep{2008EAS....29...67Z}.

Recently, {\citet{2016ApJ...833L..12V} reported observations of PSR~J1723-2837 very relevant for this study. They stated that this system is not synchronised, and inferred a ratio $P_{spin} / P_{orb} = 0.9974(7)$, where $P_{spin}$ is the rotation period of the companion and $P_{orb}$ is the orbital period of the binary. At present there are freely available observations of this object, performed by the spacecraft Kepler second mission K2 \citep{2014PASP..126..398H}. Below, we shall present our analysis, searching for the periods present in this data.}

\begin{table}
    \label{tab:Ppunto}
    \begin{threeparttable}[b]
    \caption{Change in the orbital period of different pulsars.  Parentheses are uncertainties in the last digit quoted.}
    \begin{tabular}{lr}
    \hline
    \hline
   Pulsar Name & $\dot{P}_{orb}$ [$ss^{-1}$] \\
  \hline
  $PSR~J1723-2837$\tnote{1}  & $-3.50(12) \times 10^{-9}$\\
  $2A~1822-371$\tnote{2}  & $1.51(7)\times 10^{-10}$\\
  $SAX~J17448.9-2021$\tnote{3} & $1.1(3)\times 10^{-10}$ \\
  $PSR~1957+20$\tnote{4}  & $-3.9(9)\times 10^{-11}$  \\
  $PSR~J1023+0038$ \tnote{5} & $-7.32(6)\times10^{-11}$\\
  $PSR J0740+6620$ \tnote{6} & $1.2(2)\times10^{-12}$\\
  \hline
  \hline
    \end{tabular}
    \begin{tablenotes}
    \item[1] \citet{2013ApJ...776...20C}
    \item[2] \citet{2019A&A...625L..12M} 
    \item[3] \citet{2016MNRAS.459.1340S}
    \item[4] \citet{1991ApJ...380..557R}
    \item[5] \citet{2013arXiv1311.5161A}
    \item[6] \citet{2021arXiv210400880F}
    \end{tablenotes}
    \end{threeparttable}
\end{table}

The remainder of this paper is organised as follows. In Sec.~\ref{Sec:EqsOrbEvol} we describe the tidal equations we have considered. In Sec.~\ref{sec:Results} we present a detailed analysis of our tidal model and show our results. In Sec.~\ref{sec:Asyncronew} we analyse the relevant available observations of PSR~J1723-2837 and confront them with our theoretical results.
Finally, in Sec.~\ref{sec:conclu}, we summarise the main findings of this paper, and elaborate on our conclusions.

\section{The Equations of Orbital Evolution} \label{Sec:EqsOrbEvol}

The previous calculations presented in \citet{2015ApJ...798...44B} show that the donor star in the redback system PSR~J1723-2837 is a cool object with a deep outer convective envelope. For stars with these characteristics, turbulent convection is the dominant dissipation source and it acts on the equilibrium tide \citep{2008EAS....29...67Z}. Therefore, the description of the tides we shall consider is that given by \citet{2014MNRAS.444..542R}, which is a generalisation of that given by \citet{1981A&A....99..126H}. Besides, we add the terms corresponding to gravitational wave radiation given in \citet{2002MNRAS.329..897H}.

Tides are described by a system of four ordinary, non-linear differential equations describing the evolution of the mean angular velocity $\Omega$ and eccentricity $e$ of the orbit together with the angular rotation of the companion $\omega$, and the inclination of its axis $i$ with respect to the orbital plane:
\begin{eqnarray}
 \frac{d\Omega}{dt}= 9 \bigg(\frac{K}{T}\bigg) q(1+q) \bigg(\frac{R_{2}}{a}\bigg)^8
 \frac{\Omega}{(1-e^2)^{15/2}} \nonumber \\
 \bigg[f_{1}(e^{2}) - (1-e^2)^{3/2} f_{2}(e^{2}) \frac{\omega}{\Omega} 
\cos{i} \bigg] + \bigg(\frac{d\Omega}{dt}\bigg)_{GWR},
 \label{eq:Omega_orbita}
\end{eqnarray}

\begin{eqnarray}
 \frac{de}{dt}= -27 \bigg( \frac{K}{T} \bigg) q(1+q) \bigg( \frac{R_{2}}{a} 
\bigg)^8 \frac{e}{(1-e^2)^{13/2}} \nonumber \\
 \bigg[f_{3}(e^{2}) - \frac{11}{18} (1-e^2)^{3/2} f_{4}(e^{2}) 
\frac{\omega}{\Omega} \cos{i} \bigg] + \bigg(\frac{de}{dt}\bigg)_{GWR},
 \label{eq:excentricidad}
\end{eqnarray}

\begin{eqnarray}
 \frac{d\omega}{dt}= 3 \bigg( \frac{K}{T} \bigg) \frac{q^2}{k^2} 
\bigg( \frac{R_{2}}{a} \bigg)^6 \frac{\Omega}{(1-e^2)^{6}}
 \bigg[f_{2}(e^{2})  \cos{i} - \nonumber \\ \frac{1}{4} 
\frac{\omega}{\Omega} (1-e^2)^{3/2} (3+\cos{2i}) f_{5}(e^{2}) \bigg] + \bigg(\frac{d\omega}{dt}\bigg)_{MB},
 \label{eq:rotacion1}
\end{eqnarray}

\begin{eqnarray}
 \frac{di}{dt}= -3 \bigg( \frac{K}{T} \bigg) \frac{q^2}{k^2} \bigg( 
\frac{R_{2}}{a} \bigg)^6
 \frac{\Omega}{\omega} \frac{\sin{i}}{(1-e^2)^{6}}
 \bigg[f_{2}(e^{2}) - \frac{f_{5}(e^{2})}{2}  \nonumber \\
 \bigg(\frac{\omega}{\Omega} (1-e^2)^{3/2} \cos{i} +
 \frac{R_{2}^{2}}{G M_{2}} a \omega^{2}  k^{2} (1-e^2) \bigg)
 \bigg].
\label{eq:inclinacion1}
\end{eqnarray}

There, the gravitational wave contributions \citep{2002MNRAS.329..897H} are

\begin{eqnarray}
 \frac{1}{\Omega}
 \bigg(\frac{d\Omega}{dt}\bigg)_{GWR}= -8.315 \times 10^{-10} \;
 \frac{M_{1} M_{2} M_{b} }{a^{4}}
 \frac{1+\frac{7}{8} e^{2}}{(1-e^{2})^{5/2}} \; yr^{-1},
 \label{eq:Omega_orbitaGWR}
\end{eqnarray}

\begin{eqnarray}
\frac{1}{e}
 \bigg(\frac{de}{dt}\bigg)_{GWR}= - 8.315 \times 10^{-10} \;
 \frac{M_{1} M_{2} M_{b} }{a^{4}}
 \frac{1+\frac{121}{96} e^{2}}{(1-e^{2})^{5/2}} \; yr^{-1},
 \label{eq:excentricidadGWR}
\end{eqnarray}

\noindent where $M_{1}$ and $M_{2}$ are the masses of the pulsar and the companion respectively; $M_{b}= M_{1}+M_{2}$, $q= M_{1}/M_{2}$, $k$ is the radius of gyration of the companion (the donor), which describes its moment of inertia $I$ as $I= k^{2} M_{2} R_{2}^{2}$. On the other hand, $\big(K/T\big)$ is the tidal timescale, which strongly depends on the structure of the star. The functions $f_{i}(e^{2}), i= 1, \cdots, 5$ are polynomials of the square of the eccentricity, given in \citet{1981A&A....99..126H}; for a circular orbit $f_{i}(e^2=0)= 1$.

For the magnetic braking we consider \citep{2014MNRAS.444..542R}

\begin{eqnarray}
 \bigg(\frac{d\omega}{dt}\bigg)_{MB}= -\gamma_{MB}\ R^{2}_{2}\ \omega^{3},
 \label{eq:rotacionMB}
\end{eqnarray}

\noindent where $\gamma_{MB}= 5 \times 10^{-29} s\; cm^{-2}$.
As stated above, the companion of PSR~J1723-2837 is a low mass star with a very extended outer convective zone. In this case \citep{2002MNRAS.329..897H},

\begin{equation}
 \bigg( \frac{K}{T} \bigg)=  \frac{2}{21} \frac{F_{conv}}{\tau_{conv}} 
 \frac{M_{env}}{M_{2}} yr^{-1},
\end{equation}

\noindent where $M_{env}$ is the mass in the convective envelope, $F_{conv}$ is the fraction of the convective cells which contribute to the damping, and $\tau_{conv}$ is the eddy turnover time-scale. The rate of change of the orbital period is computed as $\dot{P}_{orb}= - \big(2\pi/\Omega^2\big)\; d\Omega/dt$.

There are different proposals for expressing the coefficient $F_{conv}$. It represents the main uncertainty in the theory of tides applied to stars with convective envelopes. Here we shall analyse three different expressions for this law:

\begin{equation}
F_{conv}= {\rm min} \bigg[ 1, \bigg( \frac{P_{tid}}{2\tau_{conv}} \bigg) \bigg],
\label{f_lineal}
\end{equation}

\begin{equation}
F_{conv}= {\rm min} \bigg[ 1, \bigg( \frac{P_{tid}}{2\tau_{conv}} \bigg)^2 \bigg],
\label{f_cuad}
\end{equation}

\begin{equation}
F_{conv}= 50 \, {\rm min} \bigg[ 1, \bigg( \frac{P_{tid}}{2\tau_{conv}} \bigg)^2 \bigg],
\label{f_cuad50}
\end{equation}

\noindent where $P_{tid}$ is the tidal forcing period

\begin{equation}
\frac{1}{P_{tid}}= \bigg| \frac{1}{P_{orb}} - \frac{1}{P_{spin}} \bigg|.
\label{p_tid}
\end{equation}

\noindent Eq.~(\ref{f_lineal}) has been suggested by \citet{1966AnAp...29..489Z}, whereas Eq.~(\ref{f_cuad}) is proposed for high tidal forcing frequency (i.e., when $P_{tid} \ll \tau_{conv}$) \citep{1977Icar...30..301G}. Both have been studied under numerical simulations (\citealt{2007ApJ...655.1166P}, \citealt{2012MNRAS.422.1975O}, \citealt{2020ApJ...888L..31V}).  On another hand, Eq.~(\ref{f_cuad50}) is a result of calibrations against the cutoff period for circularization of binaries in the open cluster M67 and from the orbital decay of the high mass X-ray binary LMC X-4 \citep{2008ApJS..174..223B}.

In the tidal equations presented above it has been assumed that the donor star rotates as a rigid body. Here, we shall relax this assumption describing the donor star as a two layers object that may have different rotational velocities. It is natural to assume that these are the outer convective and inner radiative parts of the star. Magnetic braking will be coupled to the outer layers of the star, which are the ones that synchronise with the orbital period. The central part of the star acts on the outer layer contributing to modify its rotation rate. How these outer layers react will depend on the initial relative difference of rotation velocities of the two portions of the star, $\varepsilon$,

\begin{equation}
\omega_{in}=(1+\varepsilon)\; \omega_{out} ,   
\end{equation}

\noindent where $\omega_{in}$ and $\omega_{out}$ are the rotational angular velocity in the inner and outer parts of the star, respectively.
Under the condition that the inner and the outer parts synchronise their rotation for long enough times, we write for simplicity a linear coupling described by the equation

\begin{equation}
  \frac{d\omega_{in}}{dt}=\frac{I_{out}}{I \tau} (\omega_{out}-\omega_{in}) ,
   \label{Eq:rotCentral}
\end{equation}

\noindent where $\tau$ is the coupling timescale between the two stellar layers, and $I$ is the total moment of inertia, i.e, the sum of the moment of inertia of the internal ($I_{in}$) and external ($I_{out}$) parts of the star, both computed from the stellar models. Accordingly, Eq.~(\ref{eq:rotacion1}) is replaced by: 

\begin{equation}
    \frac{d\omega_{out}}{dt}= \frac{I_{in}}{I \tau} (\omega_{in}-\omega_{out})+\frac{-I(\dot{h}/I)}{I_{out}} ,
   \label{Eq:rotAfuera}
\end{equation}

\noindent where $-\dot{h}/I$ is the right hand side of Eq.~(\ref{eq:rotacion1}), being $h$ the orbital angular momentum (see \citealt{1981A&A....99..126H}). 

Summarising, the differential equations that describe our treatment of tides are (\ref{eq:Omega_orbita}),  (\ref{eq:excentricidad}), (\ref{eq:inclinacion1}), (\ref{Eq:rotCentral}), and (\ref{Eq:rotAfuera}). In Eqs.~(\ref{eq:Omega_orbita}),  (\ref{eq:excentricidad}), (\ref{eq:inclinacion1}), and in $\dot{h}$ (Eq.~\ref{Eq:rotAfuera}), we have assigned $\omega=\omega_{out}$ since we assumed that tides interact with the outer stellar layers. We have solved these equations with a fully implicit, finite differences algorithm. 

\section{Results of the model } \label{sec:Results}

\begin{figure}
\hspace*{-0.6in}
  \centering
  \includegraphics[width=0.7\textwidth]{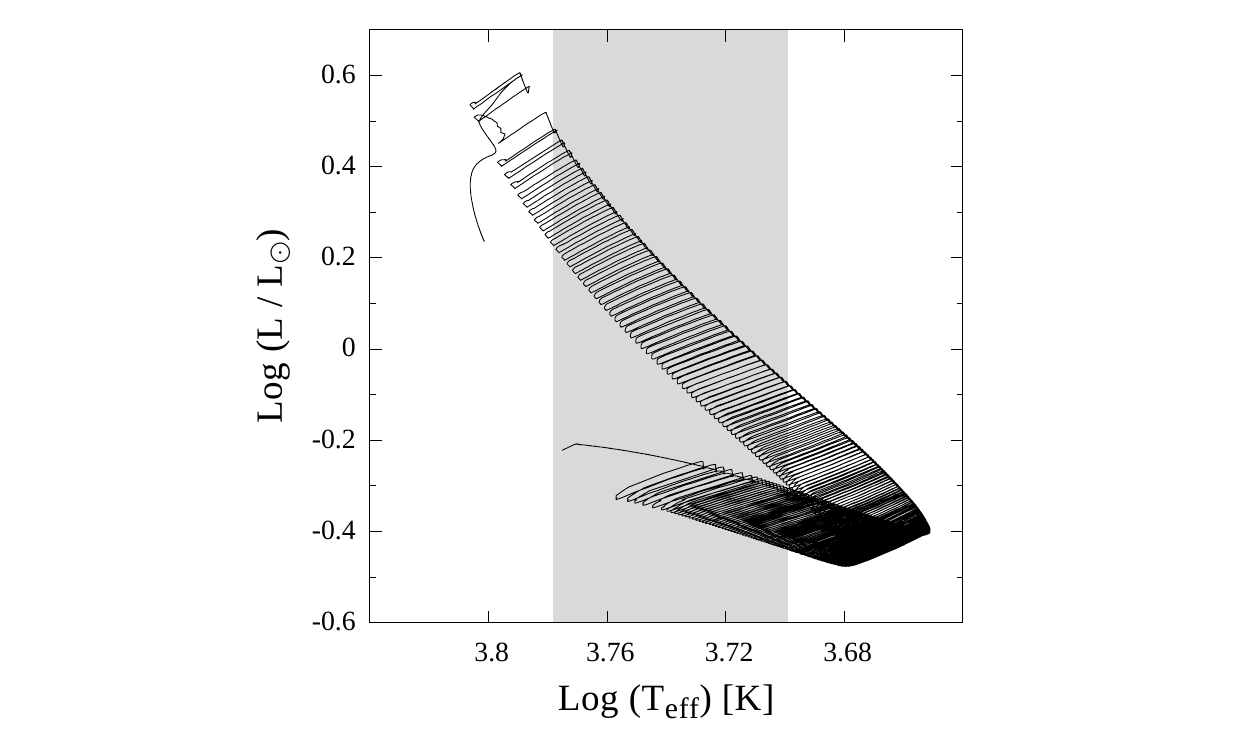}
   \caption{Evolutionary track for the donor star of a system composed by a $1.25~M_{\odot}$ solar composition donor star, evolving on a CBS together with a $1.4~M_{\odot}$ mass NS on an initial 0.75~day orbit. The calculation corresponds to an irradiation feedback regime with $\alpha_{irrad}=0.01$. The grey area depicts the range of the effective temperatures compatible with the observations of the PSR~J1723-2837 system.}
  \label{HR}
\end{figure}

\begin{figure}
\hspace*{-0.6in}
  \centering
  \includegraphics[width=0.7\textwidth]{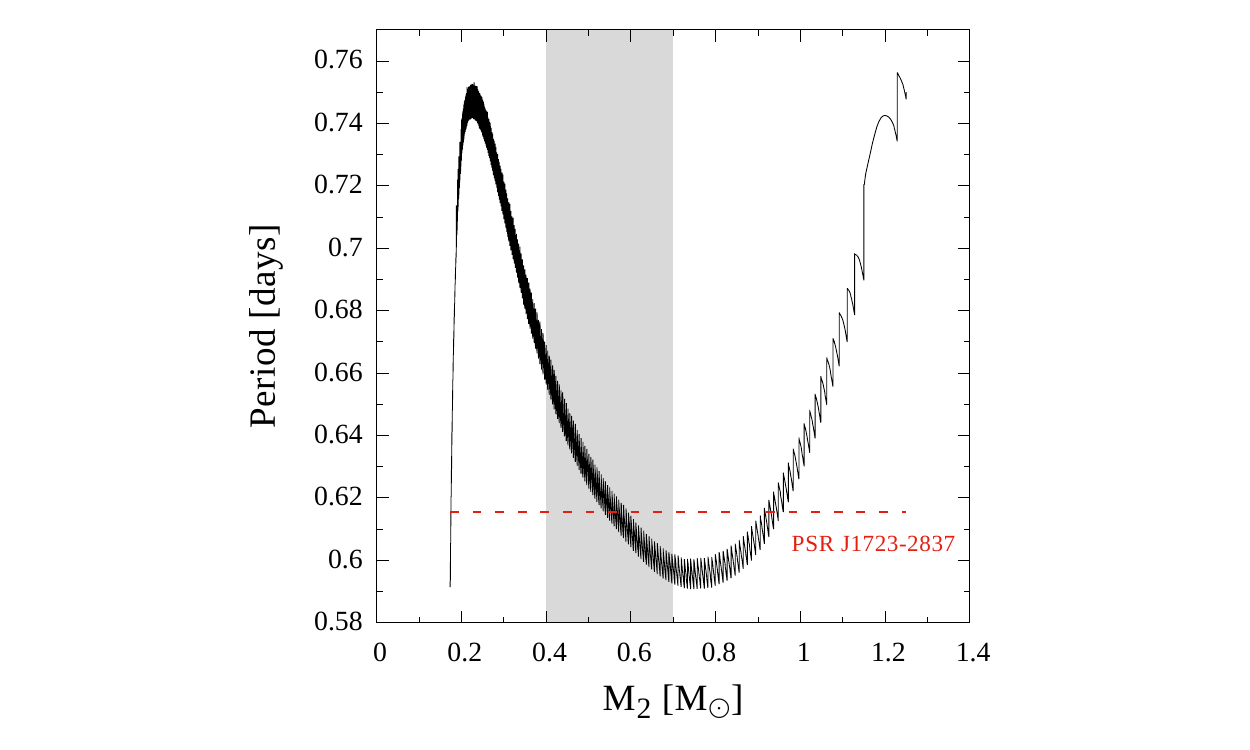}
   \caption{Orbital period as a function of the mass of the donor star for the evolutionary track presented in Fig.~\ref{HR}. The dashed line represents the orbital period observed for PSR~J1723-2837. }
  \label{PorbvsM}
\end{figure}

Using our binary evolutionary code with the inclusion of irradiation feedback (\citealt{2003MNRAS.342...50B, 2014ApJ...786L...7B, 2015MNRAS.449.4184B}), we compute the evolution of a binary that achieves the state observed for PSR J1723-2837 system. The initial parameters where taken from \citet{2015ApJ...798...44B}: a 1.25~$M_{\odot}$ solar composition donor star, evolving in a CBS together with a 1.4~$M_{\odot}$ NS on a 0.75~day orbit and an irradiation feedback regime of $\alpha_{irrad}$= 0.01. 

Fig.~\ref{HR} shows the evolutionary track in the Hertzsprung-Russell diagram for the donor star, which undergoes a large number of RLOFs, separated by detached stages. The grey area denotes the range of the  effective temperature of the donor star observed by \citet{2013ApJ...776...20C}. Fig.~\ref{PorbvsM} shows the orbital period as a function of the mass of the donor star. As it can be seen, the system achieves the observed orbital period $P_{orb}= 0.615436473(\pm8)$~d, denoted with an horizontal dashed line, during several pulses of mass transfer. In order to study the effects of tides, we select one of these pulses, i.e., a section between two successive RLOFs of the donor star, so it is in concordance with the observed orbital period and the range of temperature and mass of the donor star. It is worth to notice that any other mass transfer pulse with the same characteristics leads to similar results.

Although in this particular case the progenitor of PSR~J1723-2837 system has $\alpha_{irrad}= 0.01$, it is worth describing how a variation of this parameter would affect the tides. When $\alpha_{irrad}$ is larger, there occur less pulses, the time between pulses increases and the mass transfer rate is greater (see Fig.~1 in \citealt{2015ApJ...798...44B}) in a way in which the mass of the donor is almost unaffected. If we compare three pulses with, for example, $\alpha_{irrad}$=~1, 0.1 and 0.01 in the same moment of the evolution, luminosity, mass, orbital period and radius of the star will be very similar. The main difference will be in the change of the radius $R_{2}$ during the mass transfer pulses, because the greater the $\alpha_{irrad}$, the greater the change in $R_{2}$. Anyhow, the general behaviour would be almost the same, so qualitatively tidal interactions would bring very similar results. As stated above, for larger values of $\alpha_{irrad}$, the star remains detached longer from the Roche Lobe in between two consecutive pulses; so, tides would have more time to act. In any case, we shall show that even for the case of $\alpha_{irrad}$=~0.01, the time that the star needs to get synchronised is much smaller than the time it remains detached.

\begin{figure*}
    \centering
    \includegraphics[angle=-90,width=1.0\textwidth]{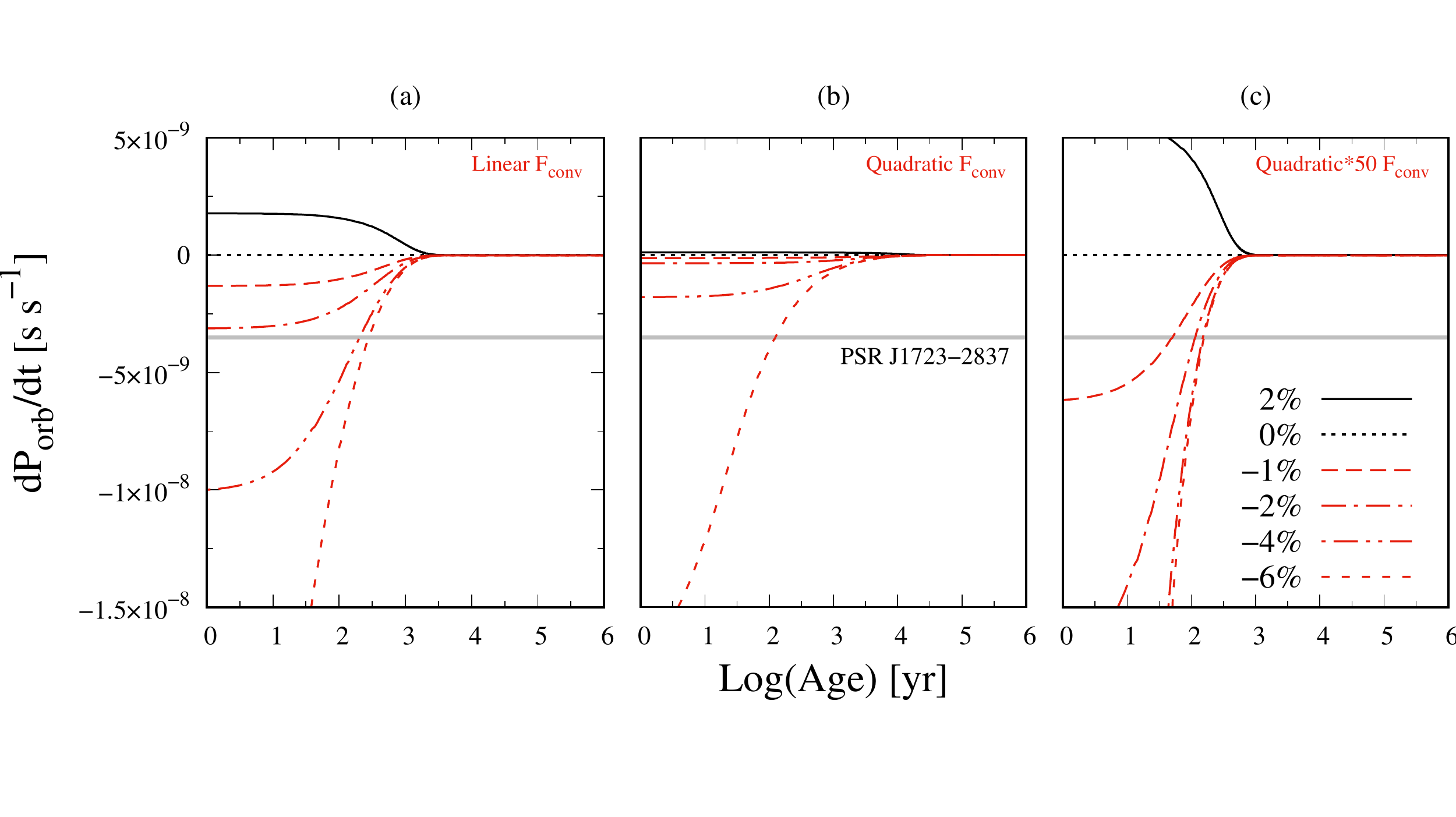}
     \caption{Change in the orbital period as a function of time for different initial asynchronism. The percentages on the right low corner of the  panel (c) give the percentage of departure from synchronism (e.g. an initial asynchronism of 2\% represents a donor in which the rotational velocity in the surface is 2\% greater than the orbital velocity). The panels correspond to each of the different expressions considered for $F_{conv}$. The horizontal grey line represents the orbital period derivative observed for PSR~J1723-2837.}
    \label{asincronismos}
\end{figure*}

\begin{figure*}
    \centering
    \includegraphics[angle=-90,width=1.0\textwidth]{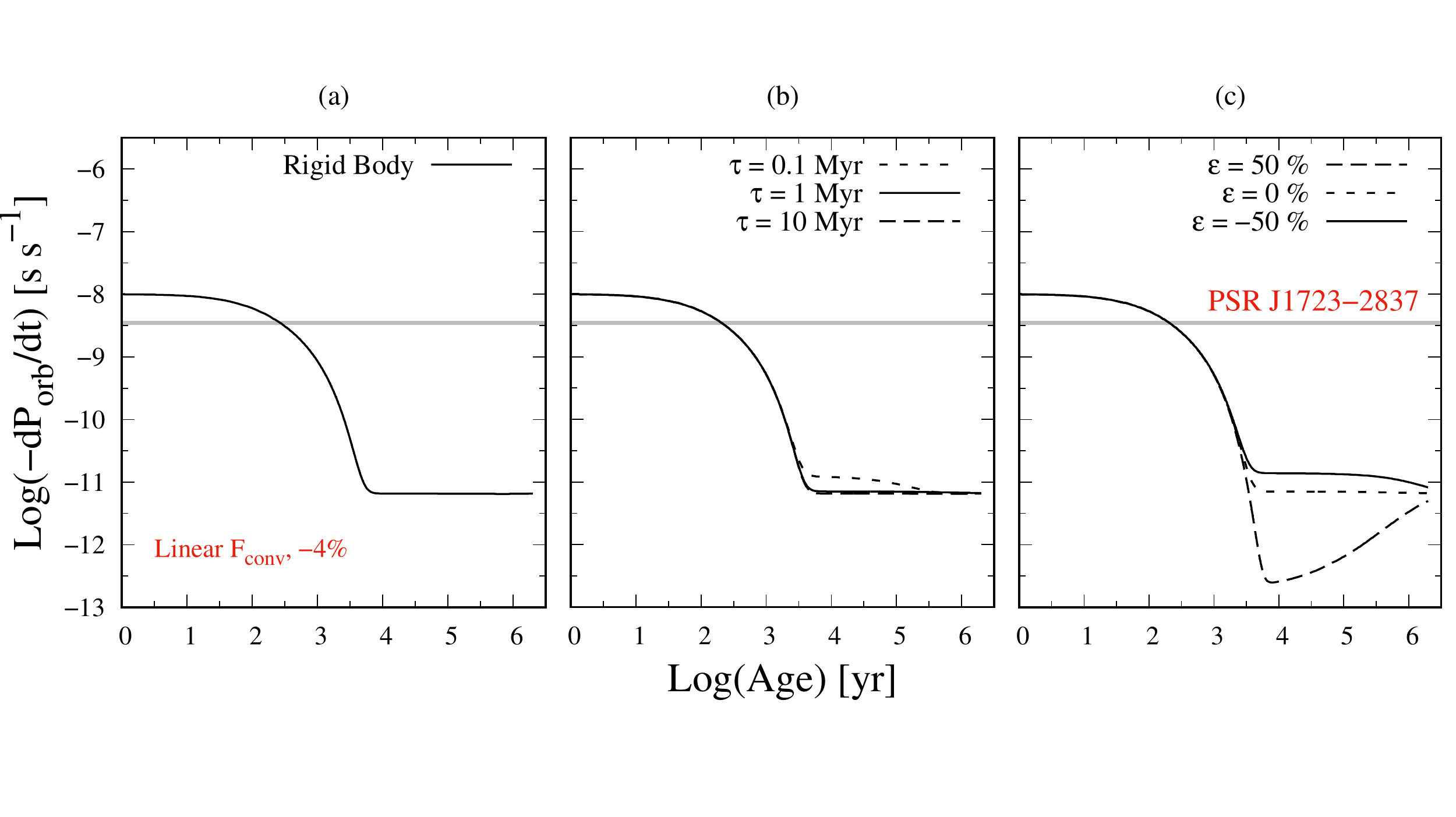}
\caption{Change in the orbital period as a function of time for a system with Linear $F_{conv}$ and an initial asynchronism of $-4\%$. Panel (a) corresponds to a star modelled as a rigid body. Panel (b) shows different values of $\varepsilon$ and $\tau$= 1~Myr. Panel (c) shows different values of $\tau$ and $\varepsilon$= 0. The horizontal grey line represents the orbital period derivative observed for PSR~J1723-2837.}
    \label{CR-epsilon-tau}
\end{figure*}

\begin{figure}
\hspace*{-0.6in}
  \centering
  \includegraphics[angle=-90,width=0.65\textwidth]{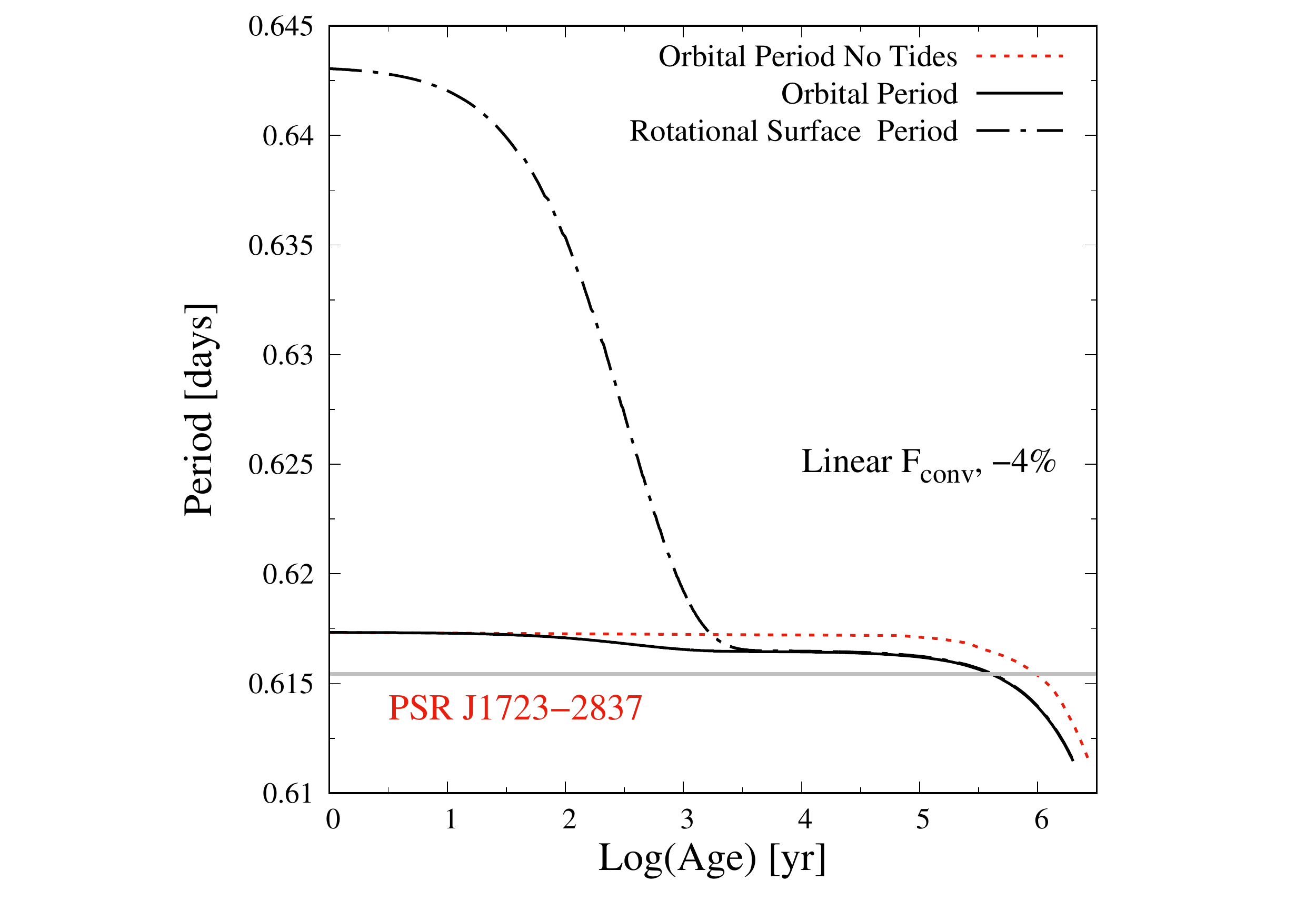}
   \caption{Orbital period and rotational period in the surface of the donor as a function of time for a system with initial asynchronism of $-4\%$, a Linear ${\rm F_{conv}}$, $\varepsilon = 0$ and $\tau = 1~Myr$. The red dotted line corresponds to the case of a rigid body without tides. The grey line represents the orbital period observed for PSR~J1723-2837.}
  \label{pp}
\end{figure}

We investigate the effects of tides on the orbital parameters in this portion of the evolution of the system, i.e., in the pulse mentioned above. For this purpose we have taken the stellar models calculated with our stellar evolutionary code and solved the full set of tidal equations presented in Sec.~\ref{Sec:EqsOrbEvol}. The results are presented in Fig.~\ref{asincronismos}, ~\ref{CR-epsilon-tau}, and ~\ref{pp} where the initial value of time is set after the detachment of the donor from its Roche Lobe, when tidal effects are computed. We consider that the system has initial values $e=0$ and $i=0$. The model of tides acting in a two layers star has three parameters: $\varepsilon$, $\tau$ (defined in Sec.~\ref{Sec:EqsOrbEvol}) and the initial asynchronism (the difference between the orbital rotational velocity and the rotational velocity in the surface of the donor star). $\varepsilon$ represents the relation between the initial rotation velocity of the two portions of the star and $\tau$ represents the characteristic coupling time between the rotation of the central region and the surface of the donor star. Typical values of $\tau$ were obtained from an independent numerical code currently under development, that solves the equations of stellar structure for differentially (shellular) rotating stars. According to our calculations, the time that it takes to a rotating layer to balance to a non-rotating one is of the order of a million years. Additionally, we explored three different laws for the scaling for the viscosity due to the turbulent convection with the tidal forcing frequency, $F_{conv}$ (see Eqs.~\ref{f_lineal}-\ref{f_cuad50}).

We study the evolution of the orbital period for each prescription of $F_{conv}$ and different initial asynchronisms, fixing $\varepsilon = 0$ and $\tau = 1~$Myr (see Fig.~\ref{asincronismos}). Each panel corresponds to a different expression for $F_{conv}$, given by Eqs.~(\ref{f_lineal}-\ref{f_cuad50}), labelled as Linear, Quadratic, and Quadratic*50 respectively. By comparing these three prescriptions, it can be seen that the initial asynchronism needed to reach the observed value of $\dot{P}_{orb}$ is very dependent on the prescription of $F_{conv}$. On the right low corner of the  panel (c) we give the percentage of departure from synchronism. For example, an initial asynchronism of 2\% represents a donor in which the rotational velocity in the surface is 2\% greater than the orbital velocity. As it can be seen from this Figure, perfectly synchronous systems (those with 0\% asynchronism) never reach the observed orbital period derivative of $-3.50(12) \times 10^{-9} ss^{-1}$. Furthermore, if the rotational velocity is greater than the orbital velocity, i.e., positive percentages of asynchronism, the change in the orbital period is positive during synchronisation. Negative values for the orbital period  derivative  are  reached only  by  systems  where  the  rotational velocity in the surface of the donor is initially smaller than the orbital velocity. This conclusion can be extended to the three laws of $F_{conv}$. Thereby, we find that, when considering the standard equilibrium theory of tidal interactions, it is possible to account for the observed value of $\dot{P}_{orb}$ if the donor star rotates a bit slower than the orbit immediately after mass transfer. 
From now on, we continue our analysis using the linear law of $F_{conv}$, since in this case, tidal forcing frequency is not high but $P_{tid}/\tau_{conv}\approx 0.2$ (see Sec.~\ref{Sec:EqsOrbEvol}). Anyhow, it is worth remarking that for the other expressions of $F_{conv}$ it is possible to find qualitatively similar results.

To explore the dependence of our models with the parameters $\varepsilon$ and $\tau$, we choose to set the initial asynchronism in $-4\%$ and then to vary these parameters. The results obtained are shown in Fig.~\ref{CR-epsilon-tau}. Panel (a) represents $\dot{P}_{orb}$ as a function of time for a star modelled as a single layer, i.e., a rigid body. Panel (b) shows the evolution of the orbital period for $\tau$ between $0.1-10$~Myr and fixed $\varepsilon = 0$. As it can be seen, varying the coupling time does not affect considerably $\dot{P}_{orb}$. Panel (c) shows a system that initially rotates as a rigid body ($\varepsilon = 0$), a system in which the outer layers of the donor star initially rotate 50\% more rapidly than the central region (positive $\varepsilon$), and a system in which the outer layers of the donor star initially rotate 50\% more slowly than the central region (negative $\varepsilon$). Here, we set $\tau = 1$~Myr. The three curves are identical for $\log{\rm(Age/yr)} \leq 3$, in the region where the observed $\dot{P}_{orb}$ is reached. After that, the curves show similar behaviour. Thus, modelling the star as a two layers object affects the value of $\dot{P}_{orb}$ when its value is far smaller (in module) than the observed value in PSR~J1723-2837.

Lastly, as can be seen in Fig.~\ref{pp}, the system synchronises in $\approx3000$~years, although the rotational period of the star remains always slightly larger than the orbital period, due to magnetic braking. The pulse of mass transfer without tides, i.e., the output of the evolution code, was chosen so it crosses the orbital period observed. Remarkably, ${P}_{orb}$ decreases more when tides act, due to the loss of energy from the system caused by this effect.

\section{Photometric analysis of PSR J1723-2837 system }
\label{sec:Asyncronew}

Recently, \citet{2016ApJ...833L..12V} have reported observations of PSR J1723--2837, claiming that the donor star is not tidally locked, having a ratio $P_{spin} / P_{orb} = 0.9974(7)$. This result was based on the detection of a series of small photometric dips, whose orbital phase seemed to decrease continuously \citep{2016MNSSA..75....9V}. These dips were interpreted as due to cold spots on the companion surface, and so the star should be rotating with a period slightly shorter than the orbital motion one.
In Section~\ref{sec:Results} we have shown that systems that are not synchronised can account for the observed orbital period derivative, but only if $P_{orb}$ is shorter than $P_{spin}$, in contrast to the result of \citet{2016ApJ...833L..12V}. Motivated by this remarkable discrepancy, we decided to make our own analysis using the photometric data obtained by the spacecraft Kepler second mission K2 \citep{2014PASP..126..398H}. The K2 data provided a precise and densely sampled light curve, suited for an accurate Fourier analysis. 

\begin{figure}
  \centering
  \includegraphics[width=0.49\textwidth]{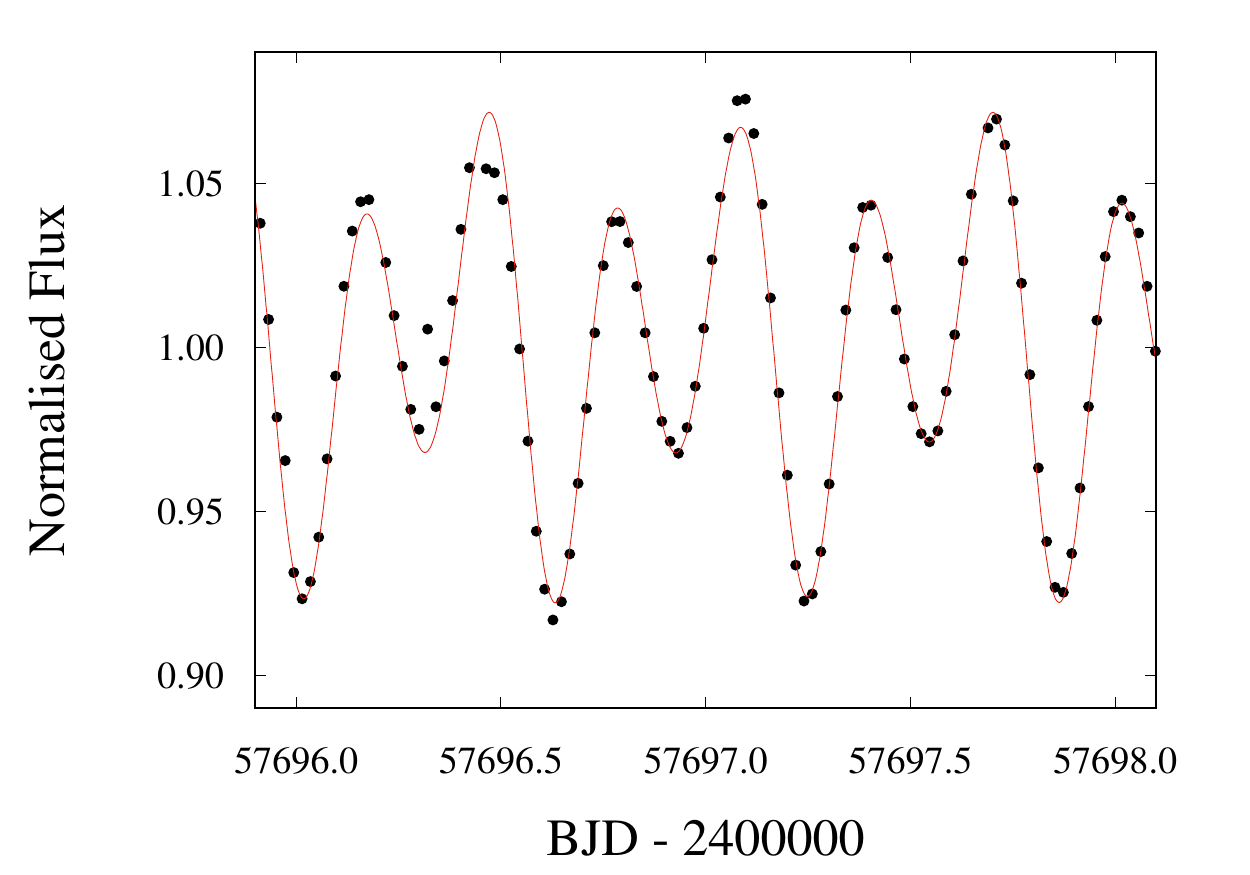}
   \caption{Portion of the unfolded light curve from long cadence K2 photometric data (black dots). 
   Two minima are observed in each cycle. The model fitted using a sine series is superimposed (solid line).}
  \label{fig:lig_cur}
\end{figure}

PSR J1723-2837 was observed by K2 (with ID 236020326) during campaign 11,  Investigation ID GO11901. High Level Science Products are publicly available for this object, in particular corrected light curves with short and long cadences (one photometric point every 60 and 1800 seconds respectively) in the wide optical spectral range characteristic of Kepler's mission \citep[see][]{2009IAUS..253..121R}.  The observations are separated in two datasets whose time spans are shown in Table \ref{tab:photom}. A small portion of the unfolded light curve, illustrating several orbital cycles, can be seen in Fig. \ref{fig:lig_cur}.

\begin{table*}
 \caption{K2 photometric datasets.}
 \label{tab:photom}
 \begin{tabular}{ccclcl}
  \hline\hline
Dataset & Cadence & \multicolumn{2}{c}{Start time}  & \multicolumn{2}{c}{End time}  \\
	     &         & Actual	   &	BJD - 2454833    & Actual &    BJD - 2454833      \\
  \hline
1 & long  & 2016-09-24 19:27:13 & 2823.3153628 & 2016-10-18 02:31:01 & 2846.6077437 \\
1 & short & 2016-09-24 19:12:59 & 2823.3054872 & 2016-10-18 02:45:14 & 2846.6176190 \\
2 & long  & 2016-10-21 06:31:48 & 2849.7746600 & 2016-12-07 23:37:45 & 2897.4825979 \\
2 & short & 2016-10-21 06:17:34 & 2849.7647845 & 2016-12-07 23:51:58 & 2897.4924732 \\
  \hline\hline
 \end{tabular}
\end{table*}

\begin{figure}
  \centering
  \includegraphics[width=0.49\textwidth]{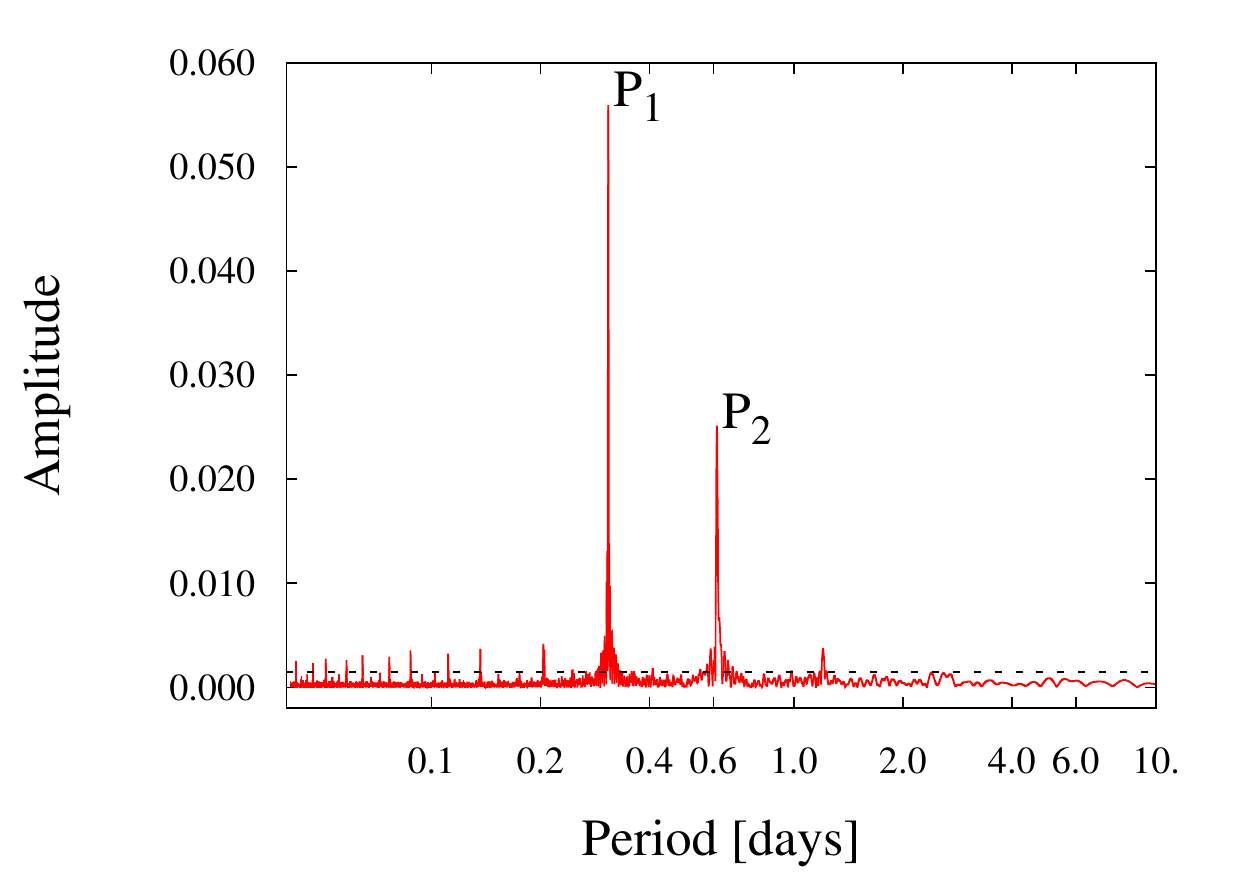}
   \caption{Amplitude spectrum of K2 photometric data. Main peaks corresponding to $P_1=0.307701$~d and $P_2=0.6129$~d are labelled. Horizontal dashed line indicate noise level.}
  \label{fig:spec_all}
\end{figure}

Aiming at identifying the periods present in the photometric oscillations, we performed a Fourier analysis of the K2 observations using the Period04 v. 1.2.0 code \citep{2004IAUS..224..786L,2005CoAst.146...53L}. The long cadence flux datasets of both campaigns were first separately normalised  by a constant and then a first Fourier spectrum of the observed fluxes was computed (Fig. \ref{fig:spec_all}). A step in frequency of $10^{-5}$~d$^{-1}$ was employed for the calculations, but it was verified that the position of the spectral peaks does not change for frequency steps one order of magnitude greater or smaller. The spectrum was calculated for frequencies between $10^{-4}$ and $24.5$~d$^{-1}$, which is approximately the Nyquist frequency for these data.  

Several prominent peaks were identified in the Fourier spectrum, corresponding the most intense to a period $\sim~0.31$~d. In order to determine its value, a series of sine functions was fitted to the data, written as
\begin{equation}
    Z + \sum A_i \sin [ 2\pi (\nu_i t + \phi_i)]
    \label{eq:sinser}
\end{equation}
 were $Z \approx 1$ is a zero point constant, $A_i$ is the amplitude, $\nu_i$ the frequency (initially set at a value corresponding to the maximum of the main peak) and $\phi_i$ the phase. For the main peak $i=1$ and the series has just one term. Then, this main sine term was subtracted from the observed data obtaining the residuals $(O-C)$. A new Fourier spectrum was then calculated from these residuals and its most intense peak ($\sim \nu_2$) identified. The original data was fitted again, this time using a two terms series, so obtaining an improved value of $\nu_1$, a first approximation of $\nu_2$ and new residuals. This process, usually known as prewhitening \citep[see Sec. 5.1.2 in][and references therein]{2010aste.book.....A}, was repeated adding each time one term to the series, until the amplitude of the most intense peak in the residuals spectrum was clearly below the noise level in the first spectrum. The parameters fitted for the first ten terms of the series are shown in Table \ref{tab:period}. The noise level estimated by Period04 for the first spectrum was $0.0015$. The amplitude of the 11th term was 0.0013 and the following  terms amplitudes, calculated until $i=17$, stabilise around $0.0012$. 
 
\begin{table}
 \caption{Periods in K2 photometric data.}
 \label{tab:period}
 \begin{tabular}{lllll}
  \hline\hline
 	&	Frequency	&	Period	&	Amplitude	&	Phase	\\
$i$	&	$\nu_i$ [d$^{-1}$]	&	$P_i$ [$d$]	&		&		\\
  \hline
1	&	3.24990(2)	&	0.307701(2)	&	0.0557(1)	&	0.1883(4)	\\
2	&	1.631(1)	&	0.6129(5)	&	0.023(3)	&	0.21(1)	\\
3	&	1.623(1)	&	0.6162(5)	&	0.016(2)	&	0.54(5)	\\
4	&	4.8749(3)	&	0.20513(1)	&	0.0034(2)	&	0.465(7)	\\
5	&	1.607(3)	&	0.622(1)	&	0.0031(8)	&	0.86(1)	\\
6	&	1.647(2)	&	0.6073(7)	&	0.003(2)	&	0.45(1)	\\
7	&	1.660(1)	&	0.6025(4)	&	0.0019(2)	&	0.22(2)	\\
8	&	0.0613(7)	&	16.3(2)   	&	0.0018(2)	&	0.30(1)	\\
9	&	0.1964(7)	&	5.09(2)   	&	0.0015(1)	&	0.79(2)	\\
10	&	0.0337(8)	&	29.7(7)  	&	0.0016(2)	&	0.50(1)	\\
  \hline\hline
 \end{tabular}
\end{table}

The detection of the main peak at $P_1 \approx P_2/2$ is due to the presence of two minima in each orbital cycle (see Fig. \ref{fig:lig_cur}). The second period detected $P_2=0.6129(5)$~d is similar to the orbital period $P_b=0.615436473(8)$~d found by \citet{2013ApJ...776...20C}, but clearly shorter, and comparable with the period proposed by \citet{2016ApJ...833L..12V} as a spin period ($P_s = 0.9974 P_b \approx 0.6138$~d).

\begin{figure}
  \centering
  \includegraphics[width=0.49\textwidth]{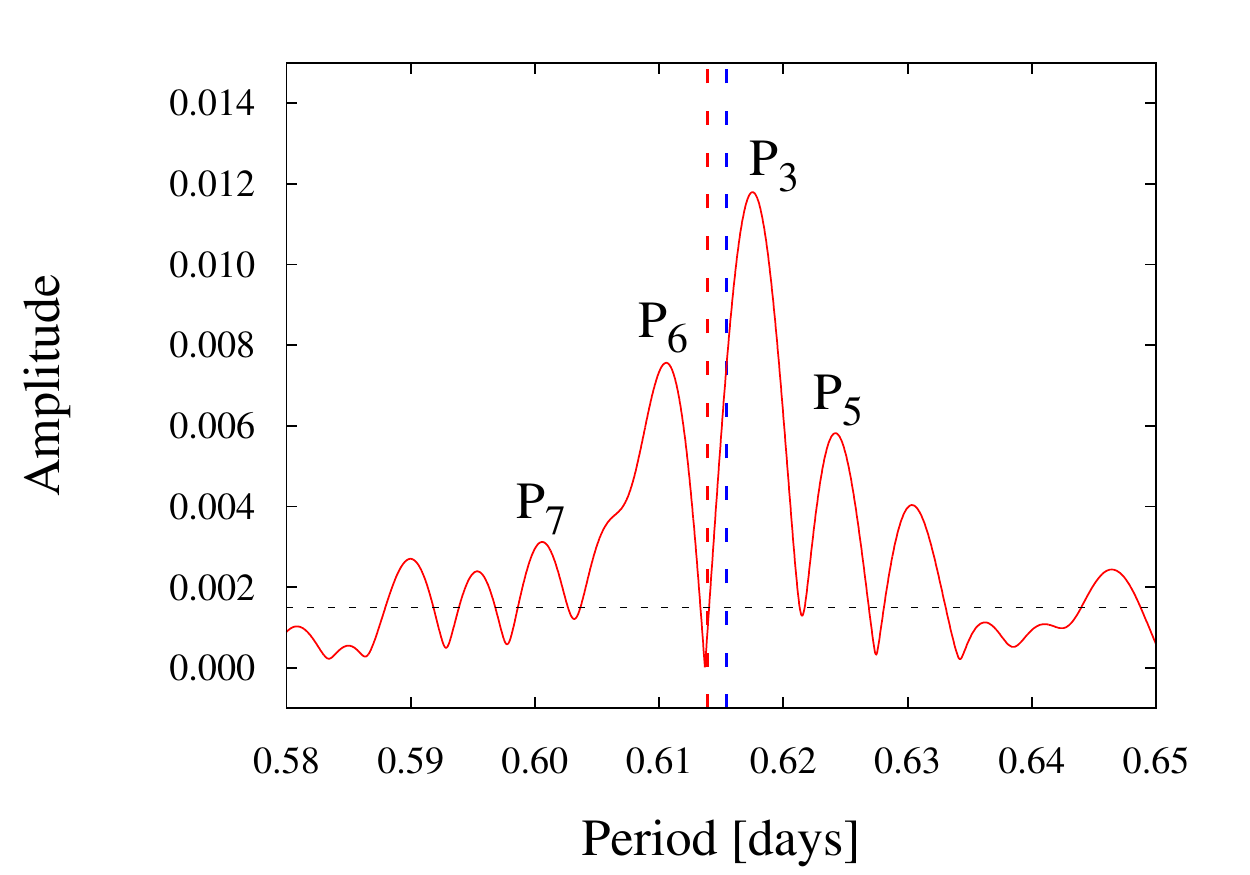}
   \caption{Fourier spectrum of residuals at original data after subtract sine terms with periods $P_1$ and $P_2$. It is indicated the position of $P_2$ peak before subtraction (red dashed line) and that of orbital period $P_b$ according to \citet{2013ApJ...776...20C} (blue). Other peaks are labelled as in Table \ref{tab:period}. Horizontal dashed line: noise level.}
  \label{fig:spec_res}
\end{figure}

However, it is worth to notice that the Fourier spectrum of the residuals at original data, after subtraction of the two firsts terms of the series (\ref{eq:sinser}), indicates the presence of at least another three periodic oscillations with periods slightly longer and shorter than $P_2$ and $P_b$, i.e. $P_3 \approx 0.6162$~d, $P_5 \approx 0.622$~d and $P_6 \approx 0.6073$~d (see Fig. \ref{fig:spec_res}). We verified that these peaks are not aliases of the main oscillation.

 As recalled above, \citet{2016ApJ...833L..12V} had already detected a photometric oscillation with a period $P_s$ shorter than the orbital one. They proposed that this oscillation is due to the presence of various spots on the stellar surface, rotating with the whole star. The new periods we found ($P_3$, $P_5$ and $P_6$) could be interpreted in the same way, i. e. as several spots on the surface of the star moving with different rotational velocities. In this case, it could be hypothesised that the star has different rotational velocities at different latitudes, something that has already been observed in other stars (e.g. the sun).

In any case, the detection of this structure around $P_2$ in the Fourier spectrum may have another interpretation. The detected frequencies so close to the rotation frequency may also be interpreted as due to the {\it beating} of non-radial pulsations perturbed by rotation. For a general description of non-radial stellar oscillations see, e.g., \citet{1989nos..book.....U,2010aste.book.....A}. In Appendix~\ref{appendix: apendA} we show a more detailed development of this idea.


\section{Discussion and Conclusions} 
\label{sec:conclu}

In a previous paper, \citet{2015ApJ...798...44B} found a progenitor to the redback system PSR~J1723-2837 that accounts for the orbital period, mass, mass ratio and temperature of the donor star, but fails to explain the orbital period derivative $\dot{P}_{orb}$. In order to continue the study of this object, we analysed the effect of tides on its orbital evolution in between two successive RLOFs of the donor star.

We based our analysis in Hut's equations for equilibrium tidal evolution in the weak friction approximation. We generalised the description of the donor star by treating it as a central sphere surrounded by a spherical layer. Accordingly, we added an extra equation that accounts for the change in the rotational velocity of the centre and added a term in the equation of the rotational velocity on the surface, which is the one that suffers the magnetic breaking effect and eventually synchronises with the orbital motion. We found two main results. One is that, considering tidal interactions on a rigid or on a two layers rotating star leads the system to similar early tidal evolution, when the observed $\dot{P}_{orb}$ value is reached. Secondly, we have shown that large negative values of $\dot{P}_{orb}$ are possible if the donor star detaches from its Roche Lobe with a rotation rate slightly lower than that of the orbit. 

This result is contrary to the effect caused by the contraction of the donor after Roche Lobe detachment. This should lead the donor to spin-up due to angular momentum conservation, and thus to faster-than-orbital rotation (if synchronization was maintained during RLOF). However, this spin-up will be small because the donor star remains in a quasi-RLOF state, as inferred by observations \citep{2013ApJ...776...20C}. In fact, according to our calculations, the filling factor $R_{2}/R_{LOBE}$ (where $R_{LOBE}$ is the radius of the Roche Lobe) is close to 0.99. Besides, only the outermost layers that are affected by irradiation are involved in the contraction, and these layers contain very little mass, so they will have a minor effect on the total angular momentum of the star. On the other hand, mass transfer will cause angular momentum loss from the donor (see, for example, \citet{2000ApJ...528..368H} for the case of isolated stars), which must be compensated by tides to maintain synchronization during RLOF. If this is not completely effective, the donor could detach from the Roche Lobe spinning somewhat slower than the orbital rate. In summary, the net effect in the variation of the angular momentum of the donor, could make possible that, when detaches, the donor star rotates somewhat slower than the whole system, as required by the tidal theory for the occurrence of $\dot{P}_{orb}<0$ values.

In brief, the theory of tidal evolution considered here indicates that systems for which the rotational velocity of the donor is slower than the orbital rotation lead to negative values of $\dot{P}_{orb}$. However, remarkably, this does not agree with the observations presented by \citet{2016ApJ...833L..12V}. This induced us to analyse photometric data obtained by the second mission K2 of the  Kepler spacecraft. We performed a Fourier analysis of the data obtaining several periods of the photometric oscillations. Two prominent peaks can be seen in the amplitude spectrum with several others less intense (see Table~\ref{tab:period}). We found several periodic oscillations in the Fourier spectrum of the residuals to the main oscillation (Fig.~\ref{fig:spec_res}), which may be associated to spots in the surface of the donor star. Furthermore, we found another possible interpretation related to non-radial oscillations of the donor star modified by its rotation.

In any case, the results presented by \citet{2016ApJ...833L..12V} and those given in our Figure~\ref{fig:spec_res} indicate the presence of few frequencies that are not easy to reconcile with the theory of tidal interactions as proposed above. This difficulty may indicate the necessity of considering tides in a more general way and that, eventually, other here overlooked phenomena may be also operating and affecting the value of $\dot{P}_{orb}$. We judge that this, as well as the origin of the presence of several frequencies close to the rotation one shown in Fig.~\ref{fig:spec_res}, as relevant findings that warrant a further study.

\section*{Acknowledgements}
We warmly thank our anonymous referee for his/her report, that helped us to largely improve the original version of this paper.

\section*{Data Availability}
The data generated by our numerical code are available from the corresponding author on request. All remaining data underlying this article are available in the article and references therein.

\bibliographystyle{mn2e}
\small
\bibliography{biblio}

\appendix
\section{Non-radial pulsations perturbed by rotation}
\label{appendix: apendA}

For pressure and gravity modes (usually referred to as p-modes and g-modes, respectively) a star has a spectrum of oscillations defined by their frequencies $\omega_{n,\ell,m}$. Here $n$ is the radial degree (related to the number of nodes in the radial part of the eigenfunction) of the oscillation; while $\ell$, the order of the mode, and $m$ (which has values in the interval $-\ell \leq m \leq \ell$) define the spherical harmonic $Y_{\ell,m}(\theta,\phi)$ that describes the angular shape of the perturbation. If the star is spherical (non rotating), $\omega_{n,\ell,m}$ is independent of $m$, i.e., it is $2\ell+1$ times degenerate. Let us denote that frequency as $\omega_{n,\ell,0}$.

It is well known that uniform rotation with angular frequency $\Omega$ fully breaks the degeneracy in $m$ to 

\begin{equation}
\omega_{n,\ell,m}= \omega_{n,\ell,0} + m\; \big(1-C_{n,\ell}\big)\; \Omega, 
\end{equation}

\noindent where $C_{n,\ell}$ is a coefficient whose value depends on the particular type of mode. The beating of two frequencies $f_{1}$ and $f_{2}$ has a frequency  $f_{beat}=|f_{1}-f_{2}|$. Immediately, (see \S~6.4 of \citealt{1989nos..book.....U}) we find that the beating of two frequencies belonging to the same $(n,\ell)$ value and $m \neq m'$ will be detected with a frequency $(\omega_{beat})_{n,\ell,m,m'}$ given by

\begin{equation}
(\omega_{beat})_{n,\ell,m,m'}= |\omega_{n,\ell,m} - \omega_{n,\ell,m'}|= |m-m'|\ \big|1-C_{n,\ell}\big|\ \Omega,
\end{equation}

\noindent which is of the order of the rotational frequency of the star. 

Regarding the splitting of mode frequencies due to rotation, it depends on the kind of the perturbed oscillation. For example, for the case of high frequency p-modes of high order $\ell$ and/or high degree $n$, $C_{n,\ell}$ is very small. In their Fig.~3.38, \citet{2010aste.book.....A} show the values of the coefficient $\beta_{n,\ell}$ for a solar model, which is very close to~1 with $|\beta_{n,\ell}-1|<0.04$; this is related to $C_{n,\ell}$ by $C_{n,\ell}= 1-\beta_{n,\ell}$. So, if $|m-m'|=1$ we shall have frequencies very close to that of rotation. If the non-radial modes have very low frequency (g-modes), rotation cannot be considered as a perturbation of the oscillation, and the problem becomes more  complicated. In any case, it seems that the beating of non-radial modes may provide an alternative, plausible explanation for the pattern shown in Fig.~\ref{fig:spec_res}. 
Of course, this is valid for a spherical star but not for a pearl-shaped object close to fill its Roche Lobe. This makes it to be a very complicate problem that it seems not have been considered to date, whose quantitative solution is far beyond the scope of the paper.


\bsp	
\label{lastpage}
\end{document}